\documentclass[english]{IEEEtran}
\usepackage{subfigure}
\usepackage[T1]{fontenc}
\usepackage[latin9]{inputenc}
\usepackage{float}
\usepackage{amsthm}
\usepackage{amsmath}
\usepackage{amssymb}
\usepackage{graphicx}
\usepackage[english]{babel}
\usepackage{epstopdf}
\usepackage{times}
\usepackage{pgfplots}
\usepackage{tikz}
\usepackage{filecontents}
\usepackage{subfig}
\usepackage{caption}
\usepackage{color}
\usepackage{balance}
\usepackage[labelsep=period]{caption}

\pgfplotsset{compat=newest} 
\pgfplotsset{plot coordinates/math parser=false} 
\newlength\figureheight 
\newlength\figurewidth 

\makeatletter

\floatstyle{ruled}
\newfloat{algorithm}{tbp}{loa}
\providecommand{\algorithmname}{Algorithm}
\floatname{algorithm}{\protect\algorithmname}

\theoremstyle{plain}

 \theoremstyle{remark}
 
\usepackage{amsfonts}
\usepackage{cite}
\usepackage{array}
\usepackage{algorithm}
\usepackage{algorithmic}
\usepackage{subfigure}
\DeclareMathOperator*{\st}{subject\ to}

\makeatother
\makeatother
\providecommand{\remarkname}{Remark}
\providecommand{\theoremname}{Theorem}

\newcommand{\bw}{\mathbf{w}}
\newcommand{\bh}{\mathbf{h}}
\newcommand{\bg}{\mathbf{g}}
\newcommand{\maxim}{\mathrm{maximize}}

\usepackage{etoolbox}

\makeatletter
\g@addto@macro\normalsize{%
 \setlength\abovedisplayskip{1.4pt}
 \setlength\belowdisplayskip{1.4pt}
 \setlength\abovedisplayshortskip{1.4pt}
 \setlength\belowdisplayshortskip{1.4pt}
}
\makeatother

\begin{document}
\title{Joint Beamforming and Antenna Selection for Sum Rate Maximization in  Cognitive Radio Networks }
\author{\IEEEauthorblockN{Van-Dinh Nguyen, \textit{Student Member, IEEE}, Chuyen T. Nguyen,\\  Hieu V. Nguyen, and Oh-Soon Shin, \textit{Member, IEEE}\vspace{-1.9\baselineskip}}
\thanks{V.-D.~Nguyen, H.~V.~Nguyen, and O.-S.~Shin are with the School of Electronic Engineering $\&$ Department of  ICMC Convergence Technology, Soongsil University, Seoul 06978, Korea  (e-mail: \{nguyenvandinh, hieuvnguyen, osshin\}@ssu.ac.kr).}
\thanks{C. T. Nguyen is with School of Electronics and Telecommunications, Hanoi University of Science and Technology, 1 Dai Co Viet, Hanoi, Vietnam (e-mail: chuyen.nguyenthanh@hust.edu.vn).}}
\maketitle
\begin{abstract}
This letter studies  joint  transmit beamforming and antenna selection at a secondary base station (BS) with multiple primary users (PUs) in an underlay cognitive radio multiple-input single-output broadcast channel. The objective is to maximize the sum rate subject to the secondary BS transmit power,  minimum   required  rates for secondary users, and PUs' interference power constraints. The utility function of interest is nonconcave and the involved constraints are  nonconvex, so this problem
is hard to solve. Nevertheless, we propose a new iterative algorithm  that finds local optima at the least. We use an inner approximation method to construct and solve  a simple convex quadratic program of moderate dimension at each iteration of the proposed algorithm.
Simulation results indicate that the proposed algorithm converges quickly  and outperforms existing approaches.
\end{abstract}
\begin{IEEEkeywords} Antenna selection, cognitive radio, nonconvex programming, sum rate, transmit beamforming. \end{IEEEkeywords}
\vspace*{-0.45cm}
\section{Introduction}
\label{introduction}
The ever-growing demand for high data rate and massive connectivity has necessitated developing advanced technologies  that can more efficiently  exploit a finite radio frequency spectrum. Among such, cognitive radio (CR)  is regarded  as a promising approach to  improve spectrum utilization \cite{Haykin05}. Specifically, primary users (PUs) in underlay CR systems have prioritized access to the available radio spectrum, and secondary users (SUs) are allowed to transmit simultaneously with PUs as long as predefined interference power constraints are satisfied at the PUs  \cite{Zhang08}.

To improve the  performance of a secondary system, the transmission strategies for SUs should be designed properly to meet a given interference power constraint. Notably, linear beamforming (BF) design  has been considered as a powerful technique that can  improve  secondary throughput. Thus, BF approaches for CR have been investigated in multiple-input single-output (MISO) \cite{HeTWC14,MaCL13} and multiple-input multiple-output (MIMO) broadcast channels \cite{Zhang09,ZhangTSP08,NguyenTVT16,NguyenCL16}.
In conjunction with BF designs, the sum rate maximization (SRM) problem of the CR network has been extensively  studied  recently. For instance, the SRM problem was investigated  with the sum power constraint (SPC) \cite{Zhang09,ZhangTSP08} and the per-antenna power constraints (PAPCs) \cite{NguyenTVT16,NguyenCL16}. However, the quality-of-service (QoS) of SUs for the SRM problem was not addressed in \cite{Zhang08,HeTWC14,Zhang09,ZhangTSP08,NguyenTVT16,NguyenCL16}, although such additional constraints are crucial to resolving the so-called user fairness. To reduce  the interference in underlay CR systems, antenna selection (AS) was proposed to select  antennas at the SUs \cite{LiTVT16} and only the best antenna at the transmitters (e.g., the source and the relay) \cite{YeohTVT14}.

In this letter, we study the SRM problem of a  CR network with constraints for  the secondary base station (BS) transmit power, SUs' minimum achievable rates, and interference power at the PUs. To mitigate the effects of the interference power constraints at the PUs which in turn improve the sum rate (SR) of the SUs, we  consider joint BF and AS (JBFAS). In addition, the proposed design  incorporates antenna selection into the power constraint to select  proper antennas at the secondary BS, differently from  \cite{LiTVT16} and \cite{YeohTVT14}.  To the authors' best knowledge, existing works cannot address the present optimization problem in
that it is difficult to even find a feasible point from a nonconvex set due to the mixed integer nature of the problem. To solve the JBFAS problem,  we propose a new iterative algorithm with low complexity. The proposed design is based on an inner approximation method that invokes  a simple convex quadratic program, which requires a lower computational effort than an exhaustive search. The obtained solutions are at least local optima since they satisfy the Karush-Kuhn-Tucker (KKT) conditions.  Numerical results show fast convergence of the proposed algorithm and a significant performance improvement over
existing  approaches.

\emph{Notation}: $\mathbf{H}^{H}$ and $\mathbf{H}^{T}$ are the Hermitian transpose and normal transpose of a matrix $\mathbf{H}$, respectively. $\|\cdot\|$ and $|\cdot|$ denote the Euclidean norm of a matrix or vector and the absolute value of a complex scalar, respectively.   $\Re\{\cdot\}$ represents the real part of a complex number. $\mathbb{E}[\cdot]$ denotes a statistical expectation. $\nabla_{\mathbf{x}}f(\mathbf{x})$ represents the gradient of  $f(\cdot)$ with respect to  $\mathbf{x}$. 
\vspace*{-0.4cm}
\section{System Model and Problem Formulation}\label{Systemmodel}\vspace*{-0.08cm}
We consider the downlink transmissions in a CR network, where a secondary BS equipped with $N_t$ transmit antennas  serves $K$ single-antenna SUs in the presence of $M$ single-antenna PUs.  It is assumed that all SUs are allowed to share the same bandwidth with the PUs for transmission \cite{ZhangTSP08,NguyenTVT16}. The channel vectors from the secondary BS to the $k$-th SU and $m$-th PU are represented by $\mathbf{h}_{k}\in\mathbb{C}^{N_t\times 1},\, k\in\mathcal{K}\triangleq\{1,2,\cdots, K\}$ and $\mathbf{g}_{m}\in\mathbb{C}^{N_t\times 1},\, m\in\mathcal{M}\triangleq\{1,2,\cdots, M\}$, respectively. We assume that  instantaneous channel state information (CSI)  is available at the transceivers for all channels, which is consistent with several
previous works on information theoretic analysis and optimization for  similar kinds of problems \cite{Zhang08,HeTWC14,Zhang09,ZhangTSP08,NguyenTVT16}. Although this assumption is quite ideal,  the assumption of perfect CSI is still of practical importance since the resulting performance serves  as a benchmark for how the CR system will perform in more realistic conditions \cite{NguyenCL16}.

The information signals are precoded at the secondary BS prior to being transmitted to the SUs. Specifically, the information intended for the $k$-th SU is $x_k\in\mathbb{C}$ with $\mathbb{E}\{|x_k|^2\} = 1$, which is precoded by beamforming vector $\mathbf{w}_k\in\mathbb{C}^{N_t\times 1}$. Then, the received signal at the $k$-th SU is given as
\begin{equation}
y_{k}=\bh^H_{k}\bw_{k}x_{k}+\sum\nolimits_{j\in\mathcal{K}\backslash \{k\}}\bh^{H}_{k}\bw_{j}x_{j}+n_{k},\label{eq:signalmodel}
\end{equation}
where $n_k\sim\mathcal{CN}(0,\sigma_k^2)$ is the additive white Gaussian noise.\footnote{Note that the background noise at the SUs also contains the interference from the primary BS, which is nonwhite in general. However, it can be assumed to be approximately white Gaussian by applying a noise-whitening filter at the SUs if the primary BS  uses a Gaussian codebook \cite{ZhangTSP08}.} Correspondingly, the achievable rate for the $k$-th SU is computed as
\begin{equation}
R_k(\bw) = \ln\Bigl(1+\frac{|\bh^H_{k}\bw_{k}|^2}{\sum_{j\in\mathcal{K}\backslash \{k\}}|\bh^{H}_{k}\bw_{j}|^2 + \sigma_k^2}\Bigr)  \label{eq:Rateformulation}
\end{equation}
where $\bw\triangleq[\bw_1^T,\cdots,\bw_K^T]^T$.

For transmit antenna selection design, let $\alpha_{n}\in\{0,1\}$ be the binary variable indicating the association of the $n$-th transmit antenna:
\begin{equation}
\alpha_n  = \left\{
							\begin{array}{ll}
									1, \mbox{ if  the $n$-th antenna is selected},   \\
									0, \mbox{ otherwise}.
							\end{array}
			   \right.
\label{eq:proba}
\end{equation}
Let us define $\tilde{\bw}_n\triangleq\bigl[[\bw_1]_n,\cdots,[\bw_K]_n\bigr]^T$ to be the beamforming weights of all  SUs associated with the $n$-th antenna, where $[\bw_k]_n$ is the $n$-th element of $\bw_k$. We impose the following constraints: 
\begin{equation}
\|\tilde{\bw}_n\|^2 \leq \alpha_n\rho_n,\,\forall n\in\mathcal{N}\triangleq\{1,\cdots,N_t\}
\end{equation}
where $\rho_n$ is a newly introduced optimization variable representing as the soft power level for the $n$-th antenna.

With the setting and explanation given above, the SRM problem based on JBFAS (JBFAS-SRM) for the CR system can be formulated as
\begin{IEEEeqnarray}{rCl} \label{eq:Sumrate:JBDAS}
& &\underset{\bw,\boldsymbol{\alpha},\boldsymbol{\rho}}{\maxim}\quad \sum\nolimits_{k=1}^{K}R_k(\bw)\IEEEyessubnumber\label{eq:JBDAS:a}\\
& &\st\; R_k(\bw) \geq \mathsf{\bar{r}}_k,\;\forall k\in\mathcal{K},\IEEEyessubnumber\label{eq:JBDAS:b}\\
& &\qquad\qquad\quad \sum\nolimits_{k=1}^K|\bg_m^H\bw_k|^2 \leq \mathcal{I}_m,\,\forall m\in\mathcal{M},\IEEEyessubnumber\label{eq:JBDAS:c}\\
 & &\qquad\qquad\quad  \|\tilde{\bw}_n\|^2 \leq \alpha_n\rho_n,\,\forall n\in\mathcal{N},\IEEEyessubnumber\label{eq:JBDAS:d}\\
& &\qquad\qquad\quad  \sum\nolimits_{n=1}^{N_t}\alpha_n\rho_n \leq P_{bs},\IEEEyessubnumber\label{eq:JBDAS:e}\\
& &\qquad\qquad\quad  \alpha_n\in\{0,1\},\,\forall n\in\mathcal{N}\IEEEyessubnumber\label{eq:JBDAS:f}
\end{IEEEeqnarray}
where $\boldsymbol{\alpha}\triangleq[\alpha_1,\cdots,\alpha_{N_t}]^T$ and $\boldsymbol{\rho}\triangleq[\rho_1,\cdots,\rho_{N_t}]^T$. 
 $\mathsf{\bar{r}}_k$ and $\mathcal{I}_m$ are the minimum required rate for the $k$-th SU and the predefined interference power at the $m$-th PU,
 respectively. $P_{bs}$ denotes the transmit power budget at the secondary BS. Note that the power constraint in \eqref{eq:JBDAS:d}-\eqref{eq:JBDAS:f} is different from SPC \cite{Zhang09,ZhangTSP08}:
\begin{IEEEeqnarray}{rCl}\label{eq:PT:SPC}
\sum\nolimits_{k=1}^{K}\|\bw_k\|^2 \leq P_{bs},
\end{IEEEeqnarray}
and PAPCs \cite{NguyenTVT16,NguyenCL16}:
\begin{IEEEeqnarray}{rCl}\label{eq:PT:PAPCs}
\sum\nolimits_{k=1}^{K}\|\bw_k\|^2_{n} \leq P_{n},\,\forall n\in\mathcal{N}
\end{IEEEeqnarray}
where $P_n$ represents the power constraint for the $n$-th antenna at the secondary BS. The antenna selection is also incorporated into the power constraint. The effect of different types of power constraints on the system performance will be discussed in  Section~\ref{Numericalresults}.
 
\section{Proposed Iterative Optimization Based Solution}\label{Proposedalgorithm}
\subsection{Relaxed Optimization Problem}
We can see that the major difficulty of solving \eqref{eq:Sumrate:JBDAS} is in finding the optimal solution for $\alpha_n$ since it is a discrete variable. Note that once $\alpha_n,\,\forall n\in\mathcal{N}$ is set to 1 or 0, the selected antennas will be fixed and thus the obtained solution may not be optimal. To circumvent this issue, we relax the constraint \eqref{eq:JBDAS:f} to $0 \leq \alpha_n\leq 1$. Consequently, the relaxed JBFAS-SRM optimization problem of \eqref{eq:Sumrate:JBDAS} can be written as  
\begin{IEEEeqnarray}{rCl} \label{eq:JBDASrelaxed}
& &\underset{\bw,\boldsymbol{\alpha},\boldsymbol{\rho}}{\maxim}\quad \sum\nolimits_{k=1}^{K}R_k(\bw)\IEEEyessubnumber\label{eq:JBDASrelaxed:a}\\
& &\st\; R_k(\bw) \geq \mathsf{\bar{r}}_k,\;\forall k\in\mathcal{K},\IEEEyessubnumber\label{eq:JBDASrelaxed:b}\\
& &\qquad\qquad\quad  \sum\nolimits_{n=1}^{N_t}\alpha_n\rho_n \leq P_{bs},\IEEEyessubnumber\label{eq:JBDASrelaxed:c}\\
& &\qquad\qquad\quad  0 \leq\alpha_n \leq 1,\,\forall n\in\mathcal{N},\IEEEyessubnumber\label{eq:JBDASrelaxed:d}\\
& &\qquad\qquad\quad \eqref{eq:JBDAS:c}, \eqref{eq:JBDAS:d}.\IEEEyessubnumber\label{eq:JBDASrelaxed:e}
\end{IEEEeqnarray}
Even with the relaxation in \eqref{eq:JBDASrelaxed:d}, the optimization problem \eqref{eq:JBDASrelaxed} is still nonconvex. Thus, it is challenging to find an optimal solution of  \eqref{eq:JBDASrelaxed}  due to the nonconcavity of the objective function and  nonconvexity of its feasible set.
In what follows, we propose an iterative algorithm that can obtain a local optimum of \eqref{eq:JBDASrelaxed}. For the set of constraints, \eqref{eq:JBDASrelaxed:b} and \eqref{eq:JBDASrelaxed:c} are nonconvex constraints while \eqref{eq:JBDASrelaxed:d} and \eqref{eq:JBDASrelaxed:e} are convex and linear constraints. Let us treat the objective function \eqref{eq:JBDASrelaxed:a} first. As observed in \cite{WES06}, \eqref{eq:Rateformulation} can be equivalently replaced by
\begin{equation}
R_k(\bw) = \ln\Bigl(1+\frac{(\Re\{\bh^H_{k}\bw_{k}\})^2}{\sum_{j\in\mathcal{K}\backslash \{k\}}|\bh^{H}_{k}\bw_{j}|^2 + \sigma_k^2}\Bigr).  \label{eq:RateRewrite}
\end{equation}
Let $\varphi_k(\bw)\triangleq\bigl(\sum_{j\in\mathcal{K}\backslash \{k\}}|\bh^{H}_{k}\bw_{j}|^2 + \sigma_k^2\bigl)/(\Re\{\bh^H_{k}\bw_{k}\})^2$, then \eqref{eq:RateRewrite} becomes $R_k(\bw) = \ln\bigl(1+1/\varphi_k(\bw)\bigl)$. Note that $R_k(\bw)$ is convex in the domain $\varphi_k(\bw) > 0$. Thus,  this is useful to develop an inner approximation of $R_k(\bw)$.    At the feasible point $\mathbf{w}^{(\kappa)}$ of $\mathbf{w}$ at the $(\kappa + 1)$-th iteration of an iterative algorithm presented shortly, a global lower bound of $R_k(\bw)$ can be obtained as \cite{Tuybook}
\begin{IEEEeqnarray}{rCl}\label{zf8:ineq}
R_k(\bw)&\geq& R_k(\bw^{(\kappa)}) + \nabla_{\varphi_k(\bw)}R_k(\bw^{(\kappa)})\bigl(\varphi_k(\bw)-\varphi_k(\bw^{(\kappa)})\bigl)\nonumber\\
&=& a_k - b_k\frac{\sum_{j\in\mathcal{K}\backslash \{k\}}|\bh^{H}_{k}\bw_{j}|^2 + \sigma_k^2}{(\Re\{\bh^H_{k}\bw_{k}\})^2} := \mathsf{R}_k^{(\kappa)}(\bw) 
\end{IEEEeqnarray}
where $a_k$ and $b_k$ are defined as
\begin{IEEEeqnarray}{rCl}
a_k &\triangleq& R_k(\bw^{(\kappa)}) + \frac{1}{1+\varphi_k(\bw^{(\kappa)})} > 0,\nonumber\\
b_k &\triangleq& \frac{1}{\varphi^2_k(\bw^{(\kappa)})+\varphi_k(\bw^{(\kappa)})} > 0.\nonumber
\end{IEEEeqnarray}
Next, at the feasible point $\mathbf{w}^{(\kappa)}$, the first-order approximation of $(\Re\{\bh^H_{k}\bw_{k}\})^2$ in \eqref{zf8:ineq} is $2\Re\{\bh^H_{k}\bw_{k}^{(\kappa)}\bh^H_{k}\bw_{k}\}-(\Re\{\bh^H_{k}\bw_{k}^{(\kappa)}\})^2$. Then, we have
\begin{IEEEeqnarray}{rCl}\label{zf8:ineq2}
\mathsf{R}_k^{(\kappa)}(\bw) &\geq &  a_k - b_k\frac{\sum_{j\in\mathcal{K}\backslash \{k\}}|\bh^{H}_{k}\bw_{j}|^2 + \sigma_k^2}{\Re\{\bh^H_{k}\bw_{k}^{(\kappa)}\}\bigl(2\Re\{\bh^H_{k}\bw_{k}\}-\Re\{\bh^H_{k}\bw_{k}^{(\kappa)}\}\bigl)}\nonumber \\
              &:=&     \widetilde{\mathsf{R}}_k^{(\kappa)}(\bw)
\end{IEEEeqnarray}
over  the trust region
\begin{equation}\label{eq:trustregion}
2\Re\{\bh^H_{k}\bw_{k}\}-\Re\{\bh^H_{k}\bw_{k}^{(\kappa)}\} > 0, \forall k\in\mathcal{K}.
\end{equation}
It should be noted that $\widetilde{\mathsf{R}}_k^{(\kappa)}(\bw)$ is concave, and \eqref{zf8:ineq} and \eqref{zf8:ineq2} are active at optimum, i.e.,
\begin{equation}\label{eq:active}
\widetilde{\mathsf{R}}_k^{(\kappa)}(\bw^{(\kappa)}) = R_k(\bw^{(\kappa)}).
\end{equation}
In order to solve \eqref{zf8:ineq2} using existing solvers such as MOSEK \cite{mosek}, we further transform  \eqref{zf8:ineq2} to 
\begin{IEEEeqnarray}{rCl}\label{zf8:ineq3}
 \widetilde{\mathsf{R}}_k^{(\kappa)}(\bw) \geq a_k - b_k\gamma_k :=\ddot{\mathsf{R}}_k^{(\kappa)}(\boldsymbol{\gamma})
\end{IEEEeqnarray}
with additional convex constraint
\begin{equation}\label{zf8:ineq4}
\frac{\sum_{j\in\mathcal{K}\backslash \{k\}}|\bh^{H}_{k}\bw_{j}|^2 + \sigma_k^2}{\Re\{\bh^H_{k}\bw_{k}^{(\kappa)}\}\bigl(2\Re\{\bh^H_{k}\bw_{k}\}-\Re\{\bh^H_{k}\bw_{k}^{(\kappa)}\}\bigl)} \leq \gamma_k,\forall k\in\mathcal{K}
\end{equation}
where $\boldsymbol{\gamma}\triangleq[\gamma_1,\cdots,\gamma_K]$ is a newly introduced variable.

Turning our attention to the constraints in \eqref{eq:JBDASrelaxed}, we see that \eqref{eq:JBDASrelaxed:b} is convex and admits the following form:
\begin{equation}\label{eq:JBDASrelaxed:b1}
 \ddot{\mathsf{R}}_k^{(\kappa)}(\boldsymbol{\gamma}) \geq \mathsf{\bar{r}}_k,\;\forall k\in\mathcal{K}.
\end{equation}
Next, for the nonconvex constraint \eqref{eq:JBDASrelaxed:c},  a convex upper bound of   $\chi_n(\alpha_n,\rho_n)\triangleq \alpha_n\rho_n$ can be found as \cite{BBA10}:
\begin{equation}\label{eq:problem:31:d4}
	 \alpha_n\rho_n\leq \frac{(\alpha_n)^2}{2r^{(\kappa)}(\alpha_n,\rho_n)}+ \frac{r^{(\kappa)}(\alpha_n,\rho_n)}{2}(\rho_n)^2  \
	                        := \chi_n^{(\kappa)}(\alpha_n,\rho_n)
\end{equation}
where  $r^{(\kappa)}(\alpha_n,\rho_n) \triangleq \alpha_n^{(\kappa)}/\rho_n^{(\kappa)} $. It is readily seen that \eqref{eq:problem:31:d4} holds with equality at optimum. Thus, \eqref{eq:JBDASrelaxed:c} is transformed to the following convex constraint:
\begin{IEEEeqnarray}{rCl}\label{eq:problem:31:d5}
	 \sum\nolimits_{n=1}^{N_t} \chi_n^{(\kappa)}(\alpha_n,\rho_n) \leq P_{bs}.
\end{IEEEeqnarray}

With the above results, we now find the solution of  \eqref{eq:JBDASrelaxed} by successively solving the following convex quadratic program $(\mathsf{JBFAS-relaxed})$:
\begin{IEEEeqnarray}{rCl} \label{eq:JBDASoptimal}
& &\underset{\bw,\boldsymbol{\alpha},\boldsymbol{\rho},\boldsymbol{\gamma}}{\maxim}\quad \sum\nolimits_{k=1}^{K}\ddot{\mathsf{R}}_k^{(\kappa)}(\boldsymbol{\gamma})\IEEEyessubnumber\label{eq:JBDASoptimal:a}\\
& &\st\quad  \eqref{eq:JBDAS:c}, \eqref{eq:JBDAS:d}, \eqref{eq:JBDASrelaxed:d}, \eqref{eq:trustregion}, \eqref{zf8:ineq4}, \eqref{eq:JBDASrelaxed:b1}, \eqref{eq:problem:31:d5}.\IEEEyessubnumber\label{eq:JBDASoptimal:b}
\end{IEEEeqnarray}
Algorithm~\ref{algo:proposedalgorithm} outlines the proposed iterative algorithm to solve the JBFAS-SRM problem \eqref{eq:JBDASrelaxed}. To generate a feasible point $(\bw^{(0)},\boldsymbol{\alpha}^{(0)},\boldsymbol{\rho}^{(0)})$ to  \eqref{eq:JBDASoptimal}, we   successively solve
\begin{IEEEeqnarray}{rCl} \label{eq:feasiblepoint}
& &\underset{\bw,\boldsymbol{\alpha},\boldsymbol{\rho},\boldsymbol{\gamma}}{\maxim}\quad \underset{k\in\mathcal{K}}{\min}\quad \{\ddot{\mathsf{R}}_k^{(\kappa)}(\boldsymbol{\gamma}) - \mathsf{\bar{r}}_k\}\IEEEyessubnumber\label{eq:feasiblepoint:a}\\
& &\st\quad  \eqref{eq:JBDAS:c}, \eqref{eq:JBDAS:d}, \eqref{eq:JBDASrelaxed:d}, \eqref{eq:trustregion}, \eqref{zf8:ineq4}, \eqref{eq:problem:31:d5}.\IEEEyessubnumber\label{eq:feasiblepoint:b}
\end{IEEEeqnarray}
until reaching $\ddot{\mathsf{R}}_k^{(\kappa)}(\boldsymbol{\gamma}) - \mathsf{\bar{r}}_k \geq 0,\;\forall k\in\mathcal{K}$.
We see that all the constraints in \eqref{eq:JBDASoptimal} are linear and quadratic. In addition, the quadratic constraints \eqref{eq:JBDAS:c}, \eqref{eq:JBDAS:d}, \eqref{zf8:ineq4}, and \eqref{eq:problem:31:d5} admit second order cone (SOC) representable \cite[Sec. 3.3]{Ben:2001}. Thus, we are able to arrive at  a SOC program, which helps  reduce the overall run-time of Algorithm~\ref{algo:proposedalgorithm}.

\begin{algorithm}[t]
\begin{algorithmic}[1]

\protect\caption{Proposed iterative algorithm for JBFAS-SRM }
\label{algo:proposedalgorithm}
\global\long\def\algorithmicrequire{\textbf{Initialization:}}
\REQUIRE  Set $\kappa:=0$ and solve \eqref{eq:feasiblepoint} to generate an initial feasible point $(\bw^{(0)},\boldsymbol{\alpha}^{(0)},\boldsymbol{\rho}^{(0)})$.
\REPEAT
\STATE Solve \eqref{eq:JBDASoptimal} to obtain the optimal solution $(\bw^{*},\boldsymbol{\alpha}^{*},\boldsymbol{\rho}^{*},\boldsymbol{\gamma}^{*})$.

\STATE Update $\mathbf{w}^{(\kappa+1)}:=\mathbf{w}^{*},$   $\boldsymbol{\alpha}^{(\kappa+1)}:=\boldsymbol{\alpha}^{*},$ and $\boldsymbol{\rho}^{(\kappa+1)}:=\boldsymbol{\rho}^{*}.$
 Set $\kappa:=\kappa+1.$
\UNTIL Convergence\\
\end{algorithmic} \end{algorithm}

\textit{Convergence analysis:} Let  $\mathcal{F}(\bw,\boldsymbol{\alpha},\boldsymbol{\rho})$ and $\mathcal{F}^{(\kappa)}(\bw,\boldsymbol{\alpha},\boldsymbol{\rho})$ denote the objective values of \eqref{eq:JBDASrelaxed} and \eqref{eq:JBDASoptimal}, respectively. It follows that 
$\mathcal{F}(\bw,\boldsymbol{\alpha},\boldsymbol{\rho})\geq
\mathcal{F}^{(\kappa)}(\bw,\boldsymbol{\alpha},\boldsymbol{\rho})$  due to \eqref{zf8:ineq}, \eqref{zf8:ineq2}, and \eqref{zf8:ineq3}.
Also, we have
$\mathcal{F}(\bw^{(\kappa)},\boldsymbol{\alpha}^{(\kappa)},\boldsymbol{\rho}^{(\kappa)})=
\mathcal{F}^{(\kappa)}(\bw^{(\kappa)},\boldsymbol{\alpha}^{(\kappa)},\boldsymbol{\rho}^{(\kappa)})$ since both \eqref{eq:active} and \eqref{zf8:ineq4} hold with equality at  optimum. This implies that Algorithm~\ref{algo:proposedalgorithm} yields a non-decreasing sequence of the
objective value, i.e., $\mathcal{F}^{(\kappa)}(\bw^{(\kappa+1)},\boldsymbol{\alpha}^{(\kappa+1)},\boldsymbol{\rho}^{(\kappa+1)}) \geq \mathcal{F}^{(\kappa)}(\bw^{(\kappa)},\boldsymbol{\alpha}^{(\kappa)},\boldsymbol{\rho}^{(\kappa)})$.
 In addition, the sequence of the objective is bounded above due to  \eqref{eq:problem:31:d5}. By following the same arguments as those in \cite[Theorem 1]{Marks:78}, we can show that Algorithm~\ref{algo:proposedalgorithm} converges to a KKT point of \eqref{eq:JBDASrelaxed}. 

\textit{Complexity analysis:} The convex problem \eqref{eq:JBDASoptimal} involves  $N_t(K+2)+K$ real scalar variables and $M+2N_t+3K+1$  quadratic and linear constraints. The computational complexity for solving \eqref{eq:JBDASoptimal} is thus $\mathcal{O}\bigl((M+2N_t+3K+1)^{2.5}\bigl((N_t(K+2)+K)^2+ M+2N_t+3K+1\bigr) \bigr)$ \cite{Ben:2001}.

\vspace*{-0.4cm}
\subsection{Improvement to Relaxed Problem}
We have numerically observed that there exists a case where $\alpha_n$ is close to 1 and $\rho_n$ is negligibly smaller than $P_{bs}$ to indicate that the $n$-th antenna is not selected. However, this value of $\rho_n$ cannot be neglected and this  will make the antenna selection procedure become  inefficient. To manage the selection exactly to in turn improve the SR, we incorporate the following additional linear constraint:
\begin{IEEEeqnarray}{rCl}\label{eq:assigningconstraint}
	 \alpha_n \leq \Omega \rho_n,\;\forall n\in\mathcal{N}	
\end{IEEEeqnarray}
where  $ \Omega $ is a given constant and is large enough to force $ \alpha_n $ to reach 0 or 1 quickly. In fact, when $\rho_n$ is  comparable to $P_{bs}$, the antenna selection satisfies
$  0 \leq \alpha_n \leq 1 \leq \Omega\rho_n,\;\forall n\in\mathcal{N},	$
to rapidly boost $\alpha_n$ up to 1. Otherwise, it is depressed to 0 by warranting $
  0 \leq \alpha_n  \leq \Omega\rho_n \leq 1,\;\forall n\in\mathcal{N},	$
when $\rho_n$ is negligibly smaller than $P_{bs}$. Here, we chose $\Omega = 100$, and this choice will not affect  the optimal solution. 

Summing up, the improved solution for the JBFAS-SRM problem in \eqref{eq:JBDASrelaxed} can be found by solving
\begin{IEEEeqnarray}{rCl} \label{eq:JBDASimproved}
& &\underset{\bw,\boldsymbol{\alpha},\boldsymbol{\rho},\boldsymbol{\gamma}}{\maxim}\quad \sum\nolimits_{k=1}^{K}\ddot{\mathsf{R}}_k^{(\kappa)}(\boldsymbol{\gamma})\IEEEyessubnumber\label{eq:JBDASimprovedl:a}\\
& &\st\;\, \alpha_n \leq \Omega \rho_n,\;\forall n\in\mathcal{N},	 \IEEEyessubnumber\label{eq:JBDASimproved:b}\\
& &\qquad\qquad\quad\eqref{eq:JBDAS:c}, \eqref{eq:JBDAS:d}, \eqref{eq:JBDASrelaxed:d}, \eqref{eq:trustregion}, \eqref{zf8:ineq4}, \eqref{eq:JBDASrelaxed:b1}, \eqref{eq:problem:31:d5}.\IEEEyessubnumber\label{eq:JBDASimproved:c}
\end{IEEEeqnarray}

\vspace*{-0.4cm}
\section{Numerical Results}
\label{Numericalresults}
In this section, we use computer simulations to investigate the performance of the proposed solution. The entries for $\mathbf{h}_k,\forall k\in\mathcal{K}$ and $\mathbf{g}_m,\forall m\in\mathcal{M}$ are generated from independent circularly-symmetric complex Gaussian random variables with zero mean and unit variance. 	It is assumed that $\sigma_k^2 =1$ for all SUs, and $P_{bs}$ and $\mathcal{I}_m,\forall m$ are defined in dB scale relative to the noise power. The  interference power constraints  for all PUs and rate constraints for all SUs are set to be equal, i.e., $\mathcal{I}_m = \mathcal{I}$, $\forall m\in\mathcal{M}$ and $\mathsf{\bar{r}}_k = \mathsf{\bar{r}}$, $\forall k\in\mathcal{K}$. We compare the system performance of the proposed design with that of the BF design using SPC in \eqref{eq:PT:SPC} and PAPCs in \eqref{eq:PT:PAPCs}. In the PAPCs design, the power constraint for each antenna is $P_n = P_{bs}/N_t$, $\forall n\in\mathcal{N}$ \cite{NguyenTVT16,NguyenCL16}.  We divide the achieved SR  by $\ln(2)$ to arrive at the unit of the bps/channel-use. 
The simulation parameters are described in the caption for ease of reference.

\begin{figure}
\centering
\includegraphics[width=0.26\textwidth,trim={1cm 0.3cm 1cm 0.5cm}]{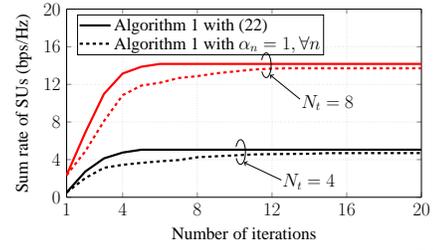}
\caption{Convergence behavior of Algorithm~\ref{algo:proposedalgorithm} ($K = M = 3$, $\mathcal{I} = 1$ dB, $\mathsf{\bar{r}} = 0.5$ bps/Hz, and $P_{bs} = 20$ dB).}
\label{fig:convergencerate}
\end{figure}

In Fig.~\ref{fig:convergencerate}, we investigate the  typical convergence behavior of Algorithm~\ref{algo:proposedalgorithm} and  also compare it with that of the case of fixing $\alpha_n = 1,\forall n$. As seen,  Algorithm~\ref{algo:proposedalgorithm}   converges very fast to the optimal value of SR, and it is insensitive to the problem size. Notably, Algorithm~\ref{algo:proposedalgorithm} with \eqref{eq:JBDASimproved} achieves a better objective value and converges faster than the case of $\alpha_n = 1,\forall n$. As expected, with a larger number of transmit antennas, we obtain a higher SR.

\begin{figure}
    \begin{center}
    \begin{subfigure}[Average SR of SUs versus $P_{bs}$ with $\mathcal{I} = 1$ dB.]{
        \includegraphics[width=0.17\textwidth,trim={2cm -0.0cm 0.1cm 0.6cm}]{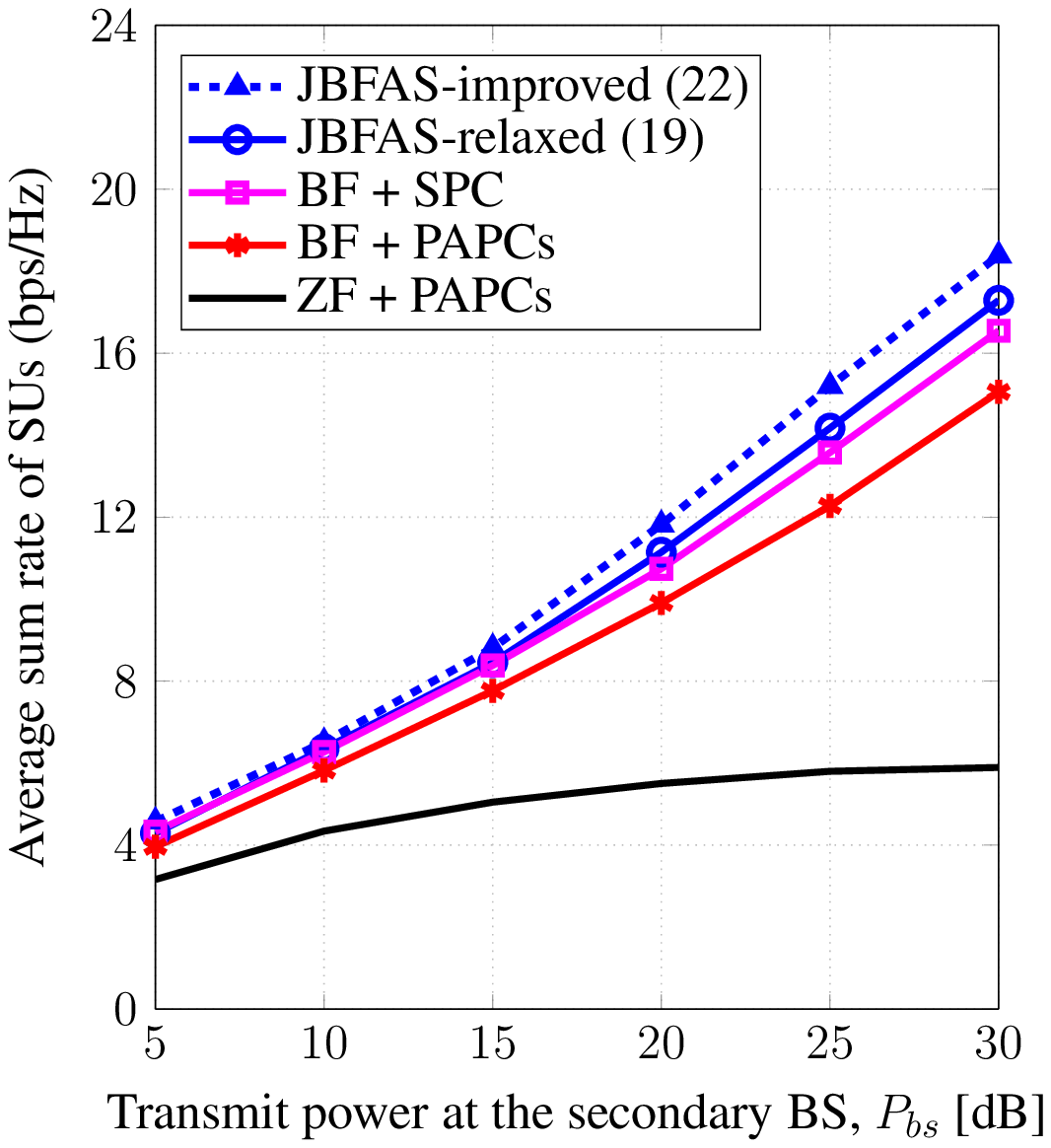}}
    		\label{fig:SRvsP}
				\end{subfigure}
		\begin{subfigure}[Average SR of SUs versus $\mathcal{I}$ with  $P_{bs} = 20 $ dB.]{
        \includegraphics[width=0.17\textwidth,trim={0.0cm -0.0cm 2cm 0.6cm}]{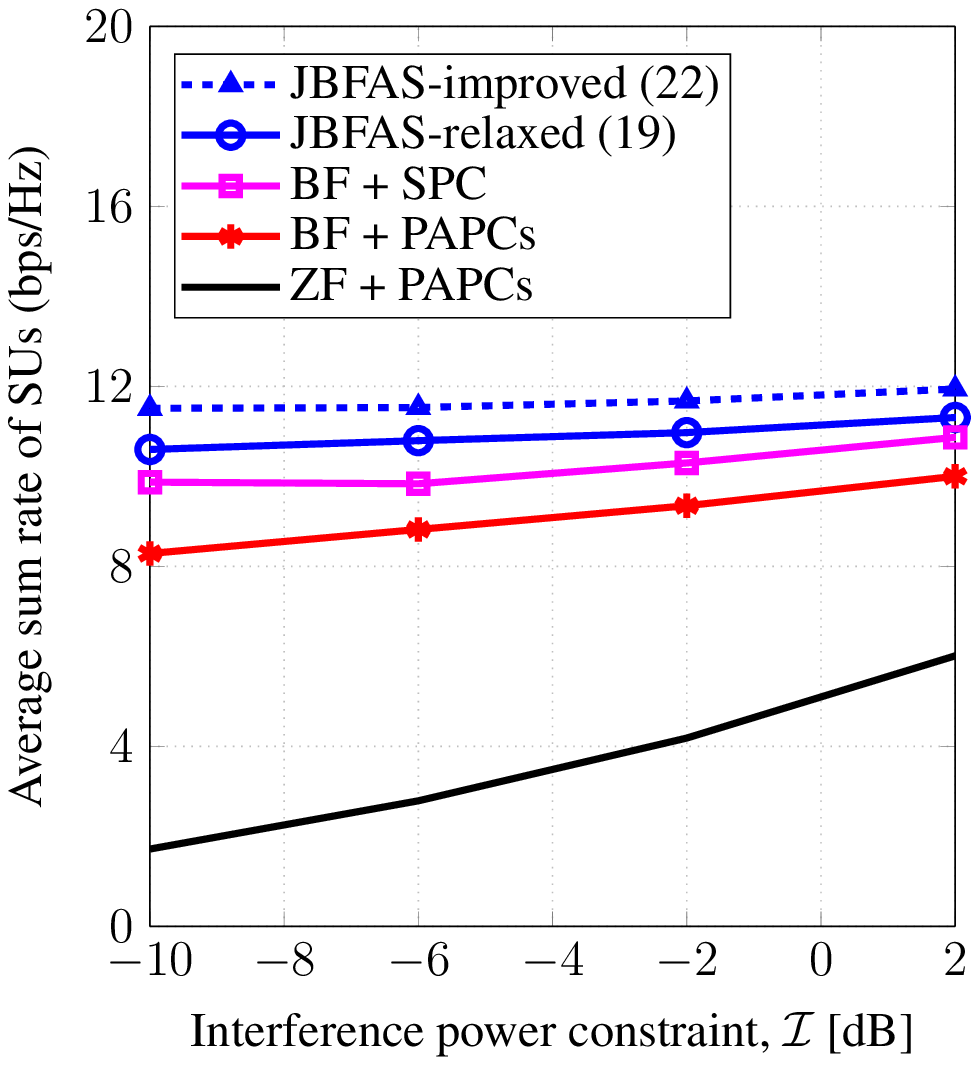}}
        \label{fig:SRvsI}
    \end{subfigure}
		\caption{Average SR of SUs, (a) versus $P_{bs}$ and (b) versus $\mathcal{I}$ ($N_t = 6, K = M = 3$, and $\mathsf{\bar{r}} = 0.5$ bps/Hz).}\label{fig:SRvsPI}
\end{center}
\end{figure}

Next, we illustrate the average SR of the SUs versus $P_{bs}$ in Fig.~\ref{fig:SRvsPI}(a) and versus $\mathcal{I}$ in Fig.~\ref{fig:SRvsPI}(b). We also compare the performance of the proposed design with zero-forcing (ZF) beamforming using PAPCs \cite{NguyenTVT16}. As can be seen, the SRs of the JBFAS designs outperform that of the others in all cases, and the gains of JBFAS designs are even higher than those of other designs for higher $P_{bs}$ (see Fig.~\ref{fig:SRvsPI}(a)) and for smaller $\mathcal{I}$ (see Fig.~\ref{fig:SRvsPI}(b)). These results are probably attributed to the fact that the secondary BS in the other designs needs to scale down its transmit power when $P_{bs}$ is high (or $\mathcal{I}$ is small) to satisfy \eqref{eq:JBDAS:c} which leads to a loss in the system performance. In contrast, only the best antennas with respect to \eqref{eq:JBDAS:c} are selected to transmit in the proposed JBFAS designs. As expected, the improved solution of \eqref{eq:JBDASimproved} achieves a larger SR compared to the relaxed solution of \eqref{eq:JBDASoptimal}.

\begin{figure}
    \begin{center}
    \begin{subfigure}[Average SR of SUs versus $M$.]{
        \includegraphics[width=0.17\textwidth,trim={2cm -0.0cm 0.1cm 0.6cm}]{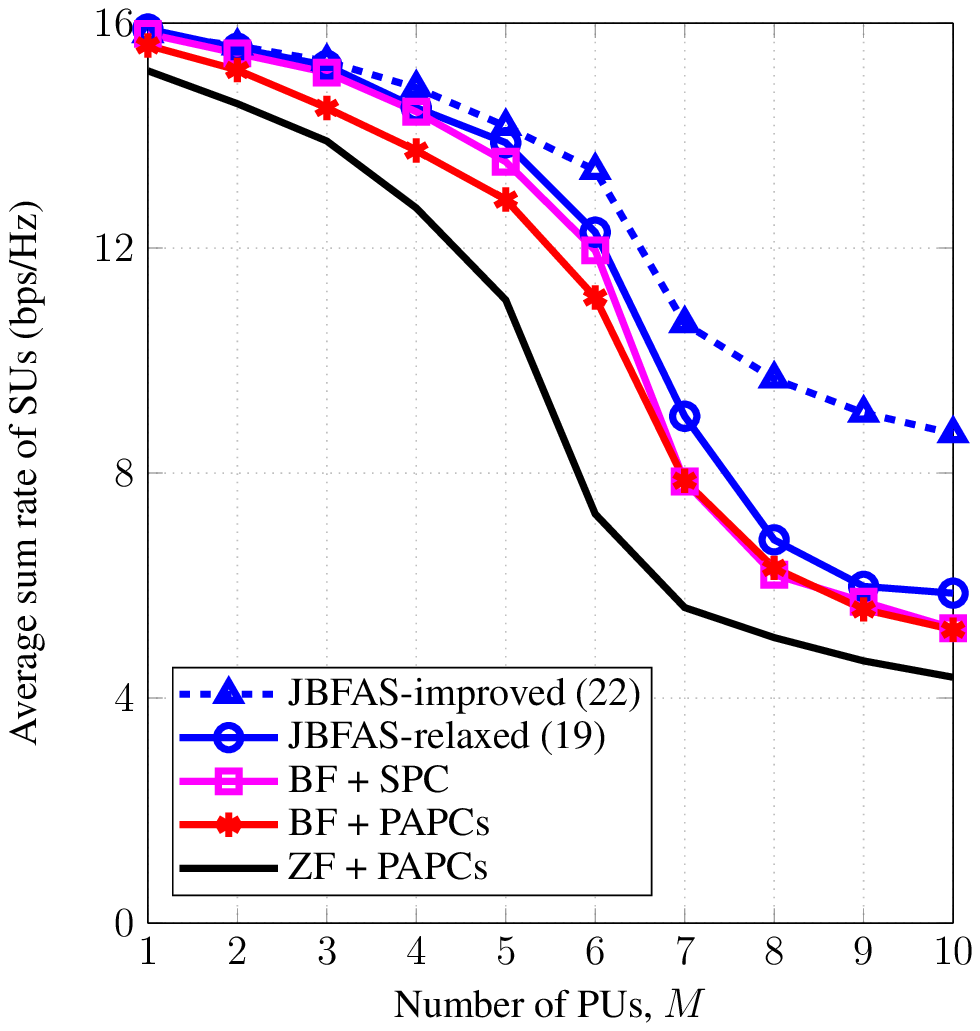}}
    		\label{fig:SRvsM}
				\end{subfigure}
		\begin{subfigure}[Average SR of SUs versus   channel uncertainties with $M=2$.]{
        \includegraphics[width=0.18\textwidth,trim={0.0cm -0.0cm 2cm 0.6cm}]{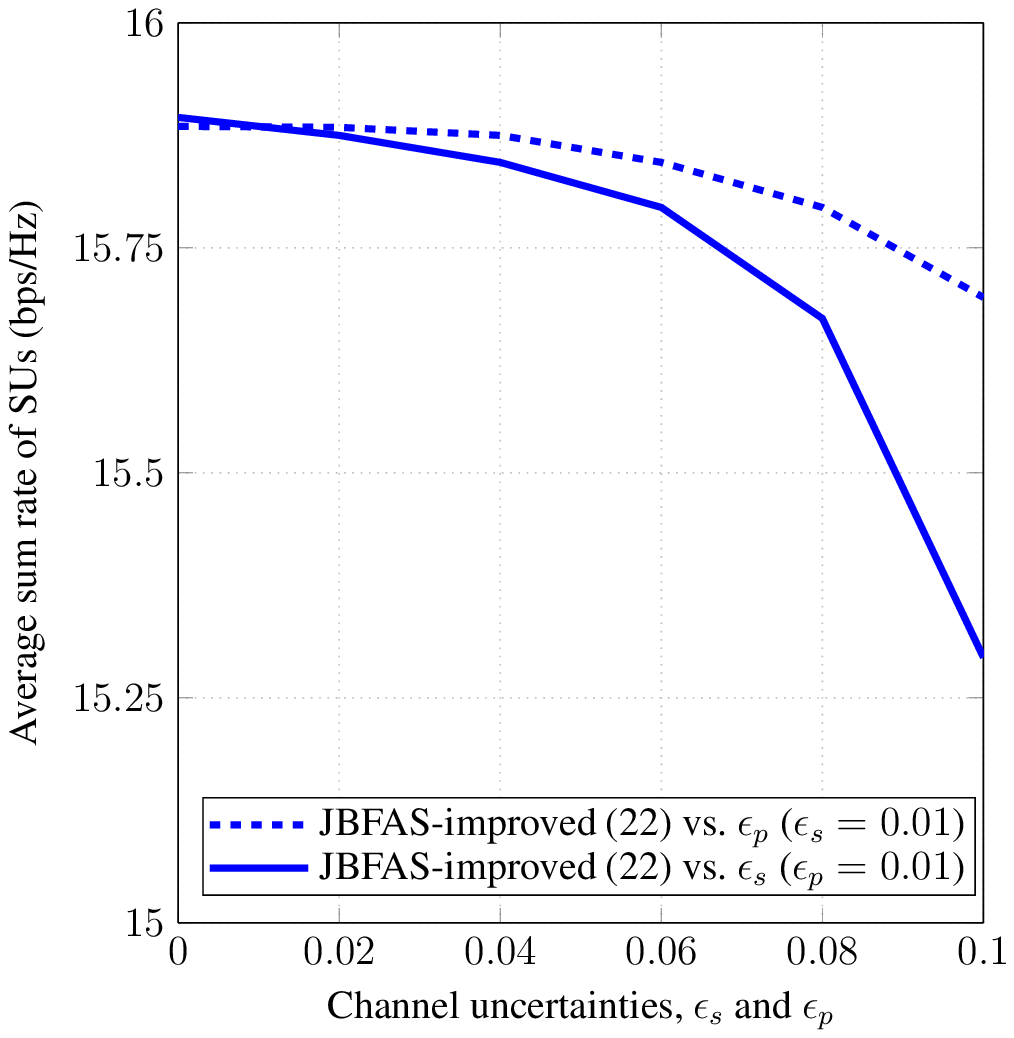}}
        \label{fig:SRvsCSI}
    \end{subfigure}
		\caption{Average SR of SUs, (a) versus $M$ and (b) versus $(\epsilon_s,\epsilon_p)$ ($N_t = 8, K = 2$, $\mathcal{I} = 2$ dB, $\mathsf{\bar{r}} = 1$ bps/Hz, and $P_{bs} = 30$ dB).}\label{fig:SRvsMCSI}
\end{center}
\end{figure}

Increasing the number of PUs drastically degrades the system performance for all designs due to a lack of the degree of freedom for
leveraging multiuser diversity, as shown in Fig.~\ref{fig:SRvsMCSI}(a). Again, the proposed JBFAS designs achieve better performance in terms of the SR compared to the others thanks to the optimized transmission.  Interestingly, the
proposed JBFAS with improved solution is quite robust to the number of  PUs.  Fig.~\ref{fig:SRvsMCSI}(b) plots   the average SR of the SUs versus different levels of channel uncertainty.  The channel vectors are modeled as $\mathbf{f} = \hat{\mathbf{f}} + \Delta\mathbf{f}$ for  $\mathbf{f}\in\{\mathbf{h}_k,\forall k,\mathbf{g}_m,\forall m\}$, where $\hat{\mathbf{f}}$ is the channel estimate available at the secondary BS and $\Delta\mathbf{f}$ represents the associated CSI error which is bounded by the uncertainty $\delta_{\mathbf{f}}$ as $\|\Delta\mathbf{f}\|^2 \leq \delta_{\mathbf{f}}$ \cite{NguyenCL16}. We define the normalized channel uncertainties as $\epsilon_s = \delta_{\mathbf{h}_k}/\|\mathbf{h}_k\|^2,\forall k$ and $\epsilon_p = \delta_{\mathbf{g}_m}/\|\mathbf{g}_m\|^2,\forall m$. From Fig.~\ref{fig:SRvsMCSI}(b), we can see that the SR is more sensitive to the CSI errors of the SUs' channels compared to those of the PUs' channels.   

\vspace*{-0.41cm}
\section{Conclusions}
\label{Conclusion}
We  presented  joint beamforming and antenna selection to maximize the SR of the secondary system of a CR network. We  developed a new iterative algorithm that quickly converges at least to a
locally optimal solution. The relaxed version of the original problem was first presented using an inner approximation method to  derive a solution. Then, we provided an improved solution by adding an additional constraint to the relaxed problem.  Numerical results  were also provided to demonstrate the  effectiveness of the proposed design. 
\vspace*{-0.4cm}

\bibliographystyle{IEEEtran}
\bibliography{Journal}
\end{document}